\documentclass[sigconf]{acmart}

\usepackage{multirow}
\usepackage{makecell}

\setcopyright{none}
\renewcommand\footnotetextcopyrightpermission[1]{} 

\AtBeginDocument{%
  }

\begin{document}

\title{A Logical-Rule Autoencoder for Interpretable Recommendations}


\author{Jinhao Pan}
\authornote{Both authors contributed equally to this research.}
\orcid{0009-0006-1574-6376}
\affiliation{%
  \institution{Department of Computer Science, George Mason University}
  \city{Fairfax}
  \state{Virginia}
  \country{USA}
  \postcode{22030}
}
\email{jpan23@gmu.edu}

\author{Bowen Wei}
\authornotemark[1]
\affiliation{%
  \institution{Department of Computer Science, George Mason University}
  \city{Fairfax}
  \state{Virginia}
  \country{USA}
  \postcode{22030}
}
\email{bwei2@gmu.edu}

\author{Ziwei Zhu}
\affiliation{%
  \institution{Department of Computer Science, George Mason University}
  \city{Fairfax}
  \state{Virginia}
  \country{USA}
  \postcode{22030}
}
\email{zzhu20@gmu.edu}


\begin{abstract}
Most deep learning recommendation models operate as black boxes, relying on latent representations that obscure their decision process. This lack of intrinsic interpretability raises concerns in applications that require transparency and accountability. In this work, we propose a Logical-rule Interpretable Autoencoder (LIA) for collaborative filtering that is interpretable by design. LIA introduces a learnable logical rule layer in which each rule neuron is equipped with a gate parameter that automatically selects between AND and OR operators during training, enabling the model to discover diverse logical patterns directly from data. To support functional completeness without doubling the input dimensionality, LIA encodes negation through the sign of connection weights, providing a parameter-efficient mechanism for expressing both positive and negated item conditions within each rule. By learning explicit, human-readable reconstruction rules, LIA allows users to directly trace the decision process behind each recommendation. Extensive experiments show that our method achieves improved recommendation performance over traditional baselines while remaining fully interpretable. Code and data are available at \url{https://github.com/weibowen555/LIA}. 
\end{abstract}

\maketitle
\pagestyle{plain} 

\section{Introduction}
Collaborative filtering (CF) is a cornerstone of modern recommender systems, and deep learning-based models have achieved remarkable performance by capturing complex user-item interactions~\cite{pan2025combating,zhu2025recommender,aljunid2025collaborative,raza2026comprehensive,zhao2024recommender,rajput2023recommender,wang2024trustworthy,torkashvand2023deep}. Despite these advances, most contemporary recommender models~\cite{koren2008factorization,sedhain2015autorec,liang2018variational} operate as black boxes, relying on latent representations and implicit neural computations. Consequently, the reasoning behind individual recommendations remains opaque. This lack of transparency is particularly problematic in high-impact applications, where trustworthiness, accountability, and fairness are essential. Without interpretable reasoning, stakeholders cannot effectively understand model behavior, diagnose errors, or justify recommendation outcomes.

In light of these concerns, prior work has explored explainable recommender systems through post-hoc techniques such as feature attribution and example-based rationales that justify recommendations by referencing similar users or items, without revealing the underlying inference mechanism~\cite{zhang2024feature,carraro2023overcoming,yao2025neural,yuan2023sequential}. Other approaches incorporate structured or symbolic components into neural architectures to enhance explainability~\cite{zhang2022neuro,chen2021neural}. However, these methods focus on generating explanations to clarify black-box outputs by analyzing feature contributions or providing approximations of model behavior, but they do not expose the actual decision logic that produces the recommendation. Consequently, the provided explanations may not faithfully reflect the actual reasoning process underlying the recommendation. As emphasized in prior work~\cite{molnar2020interpretable,rudin2019stop}, \textbf{interpretable-by-design (also called white-box) models are fully transparent and enable direct tracing of how inputs lead to predictions}. This distinction underscores the need for recommender systems that are intrinsically interpretable rather than merely explainable.


In this work, we propose a Logical-rule Interpretable Autoencoder (LIA) for collaborative filtering that is interpretable by design. At the core of LIA is a learnable logical rule layer, where each rule neuron is parameterized by a gate that automatically selects between conjunction (AND) and disjunction (OR) during end-to-end training. This learnable operator selection enables the model to discover diverse logical patterns---ranging from strict co-occurrence requirements to flexible alternative preferences---directly from user interaction data. To achieve functional completeness, the model must have the capability to express rules with negated conditions (\emph{e.g.}, a user has \emph{not} interacted with an item). LIA involves a novel negation mechanism that encodes negation directly in the sign of connection weights, providing a parameter-efficient representation that simultaneously determines item participation, polarity, and selection within each rule. Together, the learned rules form explicit, human-readable logical formulas that govern the reconstruction of user-item interactions, enabling stakeholders to precisely trace which interaction patterns contribute to a given recommendation. Experiments on three benchmark datasets demonstrate that LIA not only provides intrinsically interpretable recommendations, but also achieves improved recommendation performance over traditional baselines, without sacrificing computational efficiency.

\section{Methodology}
\begin{figure*}[t!]
    \centering
    \includegraphics[width=1.8\columnwidth]{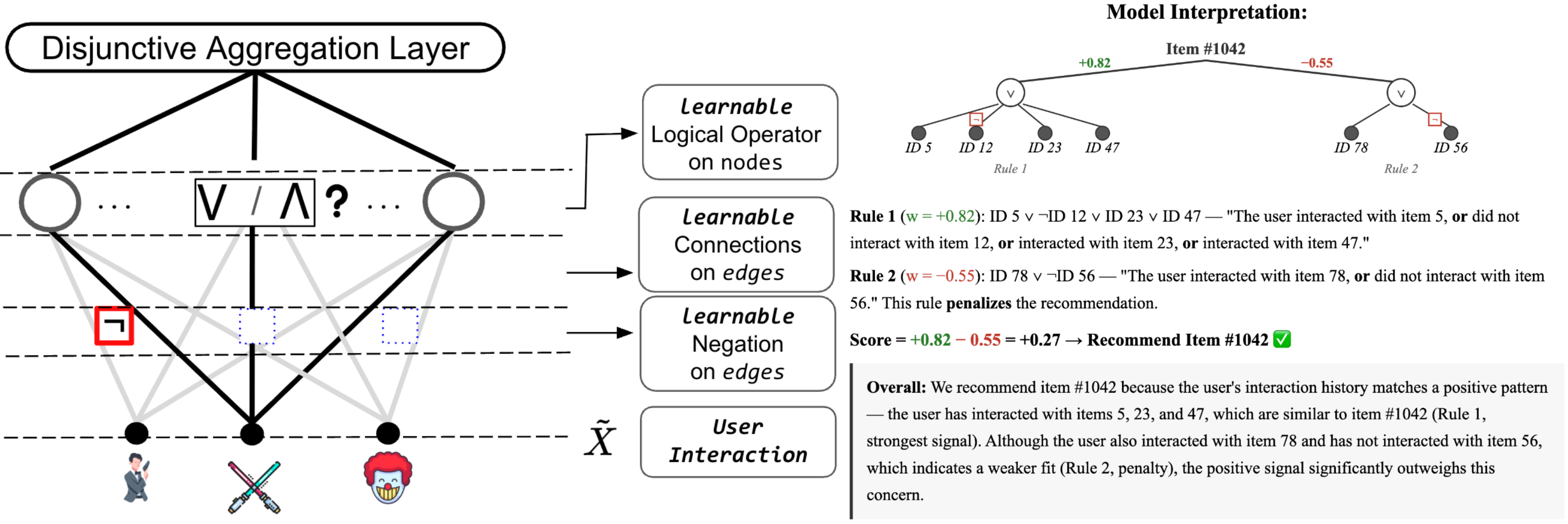}
    \vspace{-10pt}
\caption{Overview of LIA. \textbf{Left:} Architecture with signed-weight negation, learnable logical rules with operator selection, and disjunctive aggregation. \textbf{Right:} Interpretation for recommending Item \#1042, with rules visualized as trees showing learned operators, selected items, and negations (red). Natural language explanations are directly derived from the rule structure.}
    \label{fig:model}
    \vspace{-10pt}
\end{figure*}

\subsection{Problem Formulation}
Let $\mathcal{U}$ and $\mathcal{I}$ denote the sets of users and items, with $|\mathcal{U}|=M$ and $|\mathcal{I}|=N$. User-item interactions are represented by a binary matrix $\mathbf{X} \in \{0,1\}^{M \times N}$, where $x_{ui}=1$ indicates that user $u$ has interacted with item $i$. Given a user's interaction vector $\mathbf{x}_u \in \{0,1\}^N$, the goal is to produce a score vector $\hat{\mathbf{x}}_u \in \mathbb{R}^N$ that ranks unobserved items by predicted preference. In this work, we design a rule-based modeling paradigm in which logical rules over items are learned to produce recommendation scores with intrinsic interpretability.


\noindent\textbf{Interpretability via Logical Rules.} The goal is to predict a recommendation score based on human-readable logical rules. As illustrated in Figure~\ref{fig:model}, recommending Item \#1042 for User \#42 involves two rules: Rule~1 (OR over Items 5, $\neg$12, 23, 47) contributes $+0.82$ to the predicted score, while Rule~2 (OR over Items 78, $\neg$56) contributes $-0.55$ to the score, yielding a final score of $+0.27 > 0$.

\subsection{Overall Structure}
To achieve this goal, we propose LIA, a rule-based autoencoder that supports logical rule computation over items and can be trained end-to-end. We identify three key challenges: (1)~How to enable each rule to dynamically select a logical operator (AND or OR) while incorporating negation? (2)~How to support effective optimization despite the non-differentiable operations inherent in discrete logical computation? (3)~How to map rule-based representations back to item-level recommendation scores?

\noindent\textbf{Model Structure.} To address the first challenge, we design a \emph{Learnable Logical Rule Layer} where each rule neuron is parameterized by a weight vector $\mathbf{w}_k$ and a gate parameter $g_k$. The weight signs encode item negation, the weight magnitudes determine which items participate as literals, and the gate selects between AND and OR operations. To address the second challenge, we adopt a dual forward-pass mechanism with \emph{gradient grafting}~\cite{Wang_2024}: the forward pass outputs strictly binary activations for interpretability, while the backward pass routes gradients through continuous relaxations for trainability (\S\ref{sec:training}). To address the third challenge, a linear reconstruction layer maps rule activations to item scores (\S\ref{sec:recon}). The overall architecture is illustrated in Figure~\ref{fig:model}.

\noindent\textbf{Binary Rule Activations.} A key characteristic of LIA is that all rule activations are constrained to $\{0,1\}$, matching the binary nature of user--item interactions. This ensures that each rule produces a definitive True/False evaluation at inference time, enabling human-readable logical rules from the trained model.

\vspace{-10pt}
\subsection{Learnable Logical Rule Layer}
\label{sec:rule_layer}

The central component of LIA is the Learnable Logical Rule Layer, which learns $K$ logical rules over the $N$ items. We design a parameter-efficient signed-weight negation scheme and a learnable gate for automatic operator selection.

\subsubsection{Negation via Signed Weights}

To achieve functional completeness, each logical rule must be able to express negated conditions. For example, a recommendation rule might state: ``if the user interacted with item $i$ AND did \emph{not} interact with item $j$, then recommend item $k$.'' Without negation, the model can only express rules over the presence of interactions, not their absence.

A straightforward way to support negation is to double the input by concatenating $[\mathbf{x}_u;\; \mathbf{1} - \mathbf{x}_u]$, so that the network can select either $x_{ui}$ (the positive literal) or $1 - x_{ui}$ (the negated literal) for each item $i$. However, this doubles the number of parameters in every subsequent weight matrix, increasing both memory cost and the risk of overfitting.

We instead encode negation directly in the \emph{sign} of the connection weights. Each rule neuron $k$ has a weight vector $\mathbf{w}_k \in [-1, 1]^N$, where a single scalar $w_{ki}$ simultaneously determines (1)~whether item $i$ participates in the rule, and (2)~whether it appears as a positive or negated literal:
{\small
\begin{equation}
w_{ki} = \begin{cases}
> \tau & \text{positive literal (item } i \text{ present)}, \\
< -\tau & \text{negated literal (item } i \text{ absent)}, \\
\in [-\tau, \tau] & \text{item } i \text{ does not participate in rule } k,
\end{cases}
\end{equation}}
where $\tau = 0.5$ is the binarization threshold. Since we use strict inequalities ($> \tau$) in the binarization step, weights at exactly $\pm\tau$ are treated as inactive.

\subsubsection{Logical Activation Functions}

To make discrete AND/OR operations differentiable, we adopt the decoupled logical activation functions from \cite{Wang_2024}:
{\small
\begin{equation}
\mathrm{Conj}^+(\mathbf{x}_u, \mathbf{w}_k) = Q(\mathbf{1} - \mathbf{x}_u,\; \mathbf{w}_k), \quad
\mathrm{Disj}^+(\mathbf{x}_u, \mathbf{w}_k) = 1 - Q(\mathbf{x}_u,\; \mathbf{w}_k),
\end{equation}}
where $Q(\mathbf{X}, \mathbf{W}) = P(-G(\mathbf{X}) \cdot G(\mathbf{W}^\top))$, with $G(\cdot)$ and $P(\cdot)$ being element-wise functions that simulate logical operations. When inputs and weights are binary, these reduce exactly to standard AND and OR operations while remaining differentiable.
\subsubsection{Operator Selection}

Each rule $k$ is equipped with a learnable gate parameter $g_k \in \mathbb{R}$ that determines which activation function it uses. The gate selects the final binarized rule output:
{\small
\begin{equation}
\bar{r}_k = \begin{cases} \bar{r}^\wedge_k, & \text{if } g_k > 0, \\ \bar{r}^\vee_k, & \text{otherwise.} \end{cases}
\label{eq:hard_gate}
\end{equation}}
An AND rule ($g_k > 0$) fires only when \emph{all} selected literals are satisfied, while an OR rule ($g_k \leq 0$) fires when \emph{at least one} is satisfied.

In sum, each rule $k$ is fully characterized by its weight vector $\mathbf{w}_k$ (encodes both item selection and negation) and its gate $g_k$ (selects the logical operator). During training, we replace the hard binarization and gating with differentiable relaxations (Section~\ref{sec:training}) and apply the Gradient Grafting technique to maintain gradient flow.

\subsubsection{Disjunctive Aggregation Layer}
\label{sec:recon}
Stacking all $K$ rule outputs, we obtain the rule activation vector $\bar{\mathbf{r}}_u = [\bar{r}_1, \ldots, \bar{r}_K]^\top \in \{0,1\}^K$ at inference time. During training, this is replaced by its continuous relaxation $\tilde{\mathbf{r}}_u \in [0,1]^K$ to enable gradient flow (see Section~\ref{sec:training}). We use $\mathbf{r}_u$ to denote the activation vector generically.

The final prediction aggregates all $K$ rules via a disjunction: an item is recommended if \emph{any} active rule supports it. Formally, we define learnable rule--item association weights $\mathbf{W}^o \in \mathbb{R}^{N \times K}$ and bias $\mathbf{b} \in \mathbb{R}^N$, and compute item scores as: $\hat{\mathbf{x}}_u = \mathbf{W}^o \mathbf{r}_u + \mathbf{b}$. This linear combination implements a soft disjunction over rules: each column of $\mathbf{W}^o$ encodes the contribution of a rule to each item's score, and the summation aggregates evidence across all active rules. At inference time, where $\mathbf{r}_u \in \{0,1\}^K$, the score for item $j$ reduces to $\hat{x}_{u,j} = \sum_{k: \bar{r}_k=1} W^o_{jk} + b_j$, meaning that item $j$ is supported whenever at least one relevant rule fires---directly reflecting the disjunctive semantics of a logical OR.

\subsection{Training}
\label{sec:training}
The model is trained with a multinomial cross-entropy loss:
{\small
\begin{equation}
\mathcal{L} = -\frac{1}{|\mathcal{B}|}\sum_{u \in \mathcal{B}} \mathbf{x}_u^\top \log \operatorname{softmax}(\hat{\mathbf{x}}_u).
\end{equation}}
During backpropagation, gradients flow from the loss through the linear reconstruction layer to the continuous activations $\tilde{\mathbf{r}}_u$ via gradient grafting~\cite{Wang_2024}, then to the rule weights $\mathbf{w}_k$ and gate parameters $g_k$. After each gradient step, all rule weights are clipped to $[-1,1]$ to maintain valid literal selections, ensuring that the binarization threshold at $\pm 0.5$ remains meaningful throughout training.

\subsection{Rule Extraction and Interpretability.}
Each rule can be directly read off from the binarized weights and gate parameters. A dead-node elimination step prunes rules that activate for all or no users. Each surviving rule yields a human-readable logical formula. The entire reasoning chain---item selection, negation, logical operators, and rule-to-item weights---is transparent and verifiable without post-hoc approximation.

\section{Experiment}

\subsection{Setup}

\begin{table}[t!]
\caption {Statistics of three datasets.}
\vspace{-10pt}
\centering
\begin{tabular}{cccc}
\toprule
        & \#users & \#items & density \\ \midrule
ML100k    & 943   & 1,682   & 6.30\%  \\
ML1M    & 6,040  & 3,706   & 4.47\%  \\
Yelp & 12,171   & 9,252   & 0.38\%  \\ \bottomrule
\end{tabular}
\label{table:datasets}
\vspace{-15pt}
\end{table}

\subsubsection{Data and Metric}
Table~\ref{table:datasets} summarizes the statistics of the datasets used in our experiments. We evaluate our method on three public benchmark datasets: \textbf{ML100k}, \textbf{ML1M}~\cite{harper2015movielens}, and \textbf{Yelp}~\cite{yelp}. For each dataset, ratings or reviews are treated as positive user-item interactions. Interactions are randomly split into training, validation, and test sets with a ratio of 70\%, 10\%, and 20\%, respectively. We report the average NDCG@20 to evaluate the recommendation performance of different methods.

\subsubsection{Baselines}
In the experiments, we compare the proposed method with three baselines: (1) Matrix Factorization (\textbf{MF})~\cite{koren2008factorization} is the basic recommendation model that decomposes a user-item interaction matrix into latent factors with binary cross entropy loss to capture underlying patterns without any debiasing; (2) \textbf{Autoencoder}~\cite{sedhain2015autorec} is a neural recommendation model that reconstructs user-item interactions through an encoder-decoder architecture to learn latent representations of user preferences from historical interactions. (3) Multinomial Variational Autoencoder (\textbf{MultVAE})~\cite{liang2018variational} extends autoencoder-based recommendation by modeling user interaction data with a probabilistic latent variable framework and a multinomial likelihood, delivering state-of-the-art performance.


\subsubsection{Reproducibility}
\label{sec:repro}
All models are implemented in PyTorch~\cite{paszke2019pytorch} and optimized by the Adam algorithm~\cite{kingma2014adam}. All experiments are conducted on AMD CPUs and Nvidia A100-80 GB GPUs. Code and data are available at \url{https://github.com/weibowen555/LIA}.

\begin{table}[t!]
\caption{Comparing LIA with SOTA baselines on 3 datasets (NDCG@20). Best results are highlighted in \textbf{bold}.}
\vspace{-10pt}
\centering
\begin{tabular}{cccc}
\toprule
Methods  & ML100k & ML1M & Yelp \\ \midrule
MF    & .3484    & .3058  & .0747  \\
Autoencoder    & .3394   & .3102   & .0951  \\
MultVAE & .3532  & .3227  & .0953  \\ 
LIA & \textbf{.3578} & \textbf{.3263} & \textbf{.0982} \\

\bottomrule

\end{tabular}
\label{table:main_res}
\vspace{-12pt}
\end{table}

\subsection{Performance Comparison to Baselines}
We first evaluate the overall recommendation performance of LIA by comparing it with representative collaborative filtering baselines, including MF, Autoencoder, and MultVAE. All methods are evaluated on three datasets using NDCG@20. 

Table~\ref{table:main_res} shows the experimental results. LIA consistently achieves the highest NDCG@20 across all datasets, outperforming all baselines including the strongest, MultVAE. \textbf{Importantly, these gains are obtained while maintaining full interpretability}: unlike the baselines, which rely on latent representations and implicit neural computations, LIA generates recommendations through explicit, human-readable logical rules, demonstrating that competitive performance need not sacrifice interpretability.

\subsection{Efficiency Analysis}
We compare the training and inference efficiency of LIA against the baseline methods on ML1M. Table~\ref{table:efficiency} reports the training time per epoch and the inference time on the full test set.

\begin{table}[t!]
\caption{Efficiency comparison on ML1M. Training time is reported per epoch (seconds), and inference time is for the full test set (seconds).}
\vspace{-10pt}
\centering
\begin{tabular}{ccccc}
\toprule
 & MF & Autoencoder & MultVAE & LIA \\ \midrule
Train (s) & 2.42 & 0.10 & 0.05 & 0.75 \\
Infer (s) & 1.10 & 1.04 & 1.05 & 1.05 \\
\bottomrule
\end{tabular}
\label{table:efficiency}
\vspace{-15pt}
\end{table}

As shown in Table~\ref{table:efficiency}, LIA's training time per epoch (0.75s) remains moderate thanks to the efficient matrix-multiplication-based logical activation functions, and its inference time (1.05s) is comparable to all baselines. MF exhibits the longest training time due to its pairwise negative sampling procedure, while the autoencoder-based methods (Autoencoder, MultVAE) train fastest with full-batch reconstruction. At inference, LIA applies binarized rules through simple matrix multiplications without requiring iterative decoding or sampling, making it practical for real-world deployment where both efficiency and interpretability are desired.

\subsection{Hyperparameter Study}
We investigate the sensitivity of LIA to the number of rules $K$ on the ML1M dataset, which is the primary architectural hyperparameter. All other settings are fixed as described in Section~\ref{sec:repro}.

\begin{figure}[t!]
    \centering
    \includegraphics[width=0.7\columnwidth]{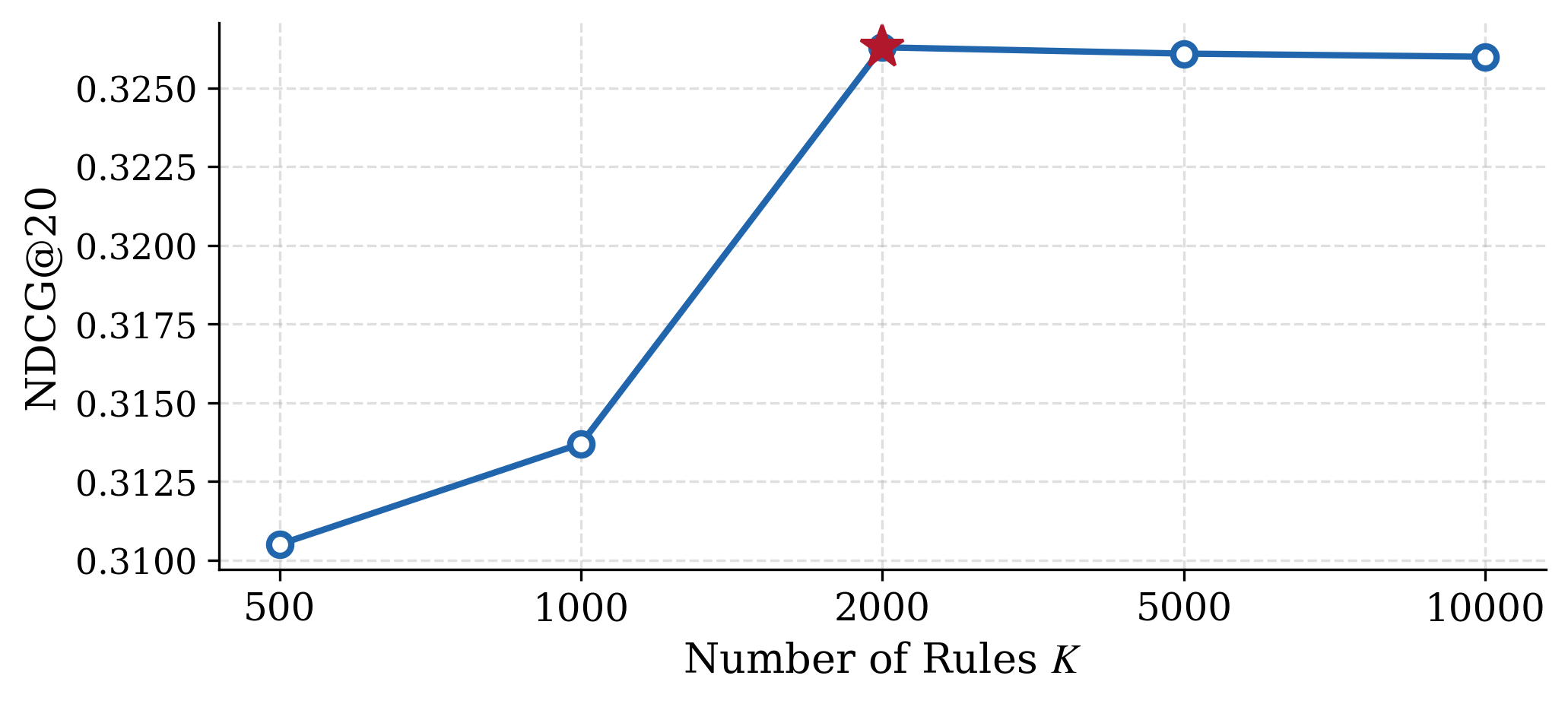}
    \vspace{-12pt}
    \caption{Effect of the number of rules $K$ on ML1M.}
    \label{fig:hyperparam}
    \vspace{-10pt}
\end{figure}

\begin{table}[t!]
\centering
\caption{Ablation study results (NDCG@20).}
\vspace{-10pt}
\label{tab:ablation}
\begin{tabular}{lccc}
\toprule
\textbf{Variant} & \textbf{ML100K} & \textbf{ML1M} & \textbf{Yelp} \\
\midrule
LIA (Full)           & .3578 & .3263 & .0982 \\
w/o Learnable Gate   & .3129 & .2718 & .0580 \\
\bottomrule
\end{tabular}
\vspace{-5pt}
\end{table}

\begin{table}[t!]
\centering
\caption{Case study: rules activated by LIA to recommend \textit{The Green Mile} for the user \#42 on ML1M.}
\vspace{-10pt}
\label{tab:case_study}
\resizebox{\columnwidth}{!}{
\begin{tabular}{clc}
\toprule
\textbf{Rule} & \textbf{Antecedent} & \textbf{Weight} \\
\midrule
AND & \textit{Shawshank} $\wedge$ \textit{Schindler's List} $\wedge$ \textit{Forrest Gump} & 0.87 \\
\addlinespace
OR  & \textit{Shawshank} $\vee$ \textit{Godfather} $\vee$ \textit{Schindler's List} $\vee$ \textit{Goodfellas} & 0.72 \\
\bottomrule
\end{tabular}
}
\vspace{-10pt}
\end{table}

\noindent\textit{Effect of the number of rules $K$.}
We vary $K \in \{500, 1000, 2000, 5000,\\ 10000\}$ and report the results in Figure~\ref{fig:hyperparam}. Performance initially improves as $K$ increases, reaching a peak at $K = 2000$, after which it plateaus or slightly degrades. A small $K$ limits the model's capacity to capture diverse user preference patterns, while a very large $K$ introduces redundant rules and increases the risk of overfitting. The result suggests that a moderate number of rules is sufficient to model the key interaction structures, which also benefits interpretability by keeping the rule set concise and human-inspectable.

\subsection{Ablation Study}
\label{sec:ablation}
To evaluate the contribution of the proposed learnable gate mechanism, we conduct an ablation study by comparing the full LIA model against a variant: \textbf{w/o Learnable Gate}, which replaces the learnable operator selection with a fixed, equal split of conjunction and disjunction rules.
The learnable gate mechanism proves to be the most critical component of LIA. Removing it causes substantial performance drops of 12.5\%, 16.7\%, and 40.9\% on ML100K, ML1M, and Yelp, respectively. Without the learnable gate, the model falls back to a fixed allocation of conjunction and disjunction operators, which cannot adapt the logical operator type to each rule's needs. This confirms that allowing each rule to select its logical operator autonomously is essential for capturing diverse user preferences.

\subsection{Case Study}

To qualitatively demonstrate the interpretability of LIA, we present a case study from the ML1M dataset in Table~\ref{tab:case_study}. Consider User~\#42, who has interacted with \textit{The Shawshank Redemption}, \textit{Schindler's List}, and \textit{Forrest Gump}. LIA recommends \textit{The Green Mile} by activating two interpretable rules shown in Table~\ref{tab:case_study}.

The AND rule captures a conjunctive co-occurrence: users who watched all three 90s prestige dramas are highly likely to watch \textit{The Green Mile}. The OR rule captures a broader thematic cluster of critically acclaimed dramas: interacting with \emph{any} of these suffices to activate the rule. Together, these rules show that LIA learns patterns at varying granularity. Importantly, these rules are not post-hoc rationalizations but integral components of the prediction---each rule's weight directly contributes to the final score, ensuring faithful interpretations. In contrast, baselines (e.g., MF and MultVAE) produce the same recommendation through opaque latent computations, offering no insight into \emph{why} the item was suggested.

\section{Related Work}
\noindent Explainable recommendation models seek to provide human understandable justifications for recommendation results without necessarily exposing the internal inference process of the recommender. Existing approaches can be broadly categorized into three types. (1) Post-hoc explanation methods generate explanations after predictions are made by analyzing black-box models~\cite{zhang2014explicit,tan2021counterfactual,zhong2022shap}, for example through feature attribution, attention visualization, or example-based rationales; these explanations approximate model behavior but are not guaranteed to reflect its true decision logic. (2) Textual and aspect-based explanation models leverage auxiliary textual data such as user reviews or item descriptions to produce natural-language explanations that highlight preference aspects (e.g., ``battery life'')~\cite{zhang2020explainable,li2021personalized,tai2021user,xian2020cafe}, yet the recommendation inference itself remains opaque. (3) Attribute-based and rule-oriented models incorporate explicit user or item attributes to generate structured rationales~\cite{zhang2024feature,yao2025neural,yuan2023sequential,tal2019neural}, including neuro-symbolic approaches that learn attribute-grounded rules to justify recommendations~\cite{zhang2022neuro,chen2021neural,carraro2023overcoming}. While these methods improve transparency at the explanation level, they primarily explain recommendation outcomes rather than expose an intrinsically transparent reasoning process. As a result, existing explainable recommender systems do not achieve full end-to-end interpretability.

Interpretable-by-design recommender systems are designed such that the reasoning behind each recommendation can be directly followed and verified by humans. Classical neighborhood-based methods, such as k-nearest neighbors (kNN) and user/item-based collaborative filtering, exhibit a degree of interpretability by relying on explicit similarity computations among users or items~\cite{ahuja2019movie,singh2020movie,he2017neural,sarwar2001item}. However, these approaches are limited in expressiveness and struggle to capture complex, high-order user preferences, leading to inferior performance compared to modern neural recommenders. Beyond such heuristic methods, learning-based recommender models are typically non-interpretable by design, as their inference relies on latent neural representations. Consequently, accurate and intrinsically interpretable recommender systems remain scarce. We address this gap with an interpretable collaborative filtering model that provides transparent, human-inspectable reasoning.

\section{Conclusion}
In this work, we examined the problem of interpretability in collaborative filtering and argued that most existing recommender models, while effective, lack intrinsic transparency due to their reliance on latent representations and implicit neural computations. To address this limitation, we proposed LIA, a logical-rule interpretable autoencoder for collaborative filtering that provides intrinsic interpretability through explicit rule-based inference. Extensive experiments showed that LIA achieves improved recommendation performance over traditional baselines while preserving transparency and efficiency. This work demonstrates that accurate recommendation and interpretable reasoning can be jointly achieved, and highlights the potential of rule-based learning paradigms for building trustworthy recommender systems.


\bibliographystyle{ACM-Reference-Format}
\bibliography{paper}


\end{document}